\documentstyle[sprocl]{article}
\def\arcmin{\hbox{$^\prime$}}
\def\farcm{\hbox{$.\!\!^{\prime}$}}
\def\farcs{\hbox{$.\!\!^{\prime\prime}$}}

\input{psfig}
\begin{document}

\title{Constraining halo properties from galaxy-galaxy lensing and photo-z}
\author{Andreas O. Jaunsen}
\address{European Southern Observatory, Casillo 19001, Santiago 19, Chile \\ E-mail: ajaunsen@eso.org}

\maketitle
\abstract{
Here we present results from a maximum likelihood analysis of galaxy-galaxy
weak lensing effects as measured in a $12\farcm5 \times 12\farcm5$ field
obtained at the Nordic Optical Telescope, on La Palma, Spain. The analysis
incorporates photometric redshifts and gives circular velocities consistent
with previous weak lensing work.
}

\section{Data}

The Canada France Redshift Survey (CFRS) 14h-field was observed in May
1998 with the Nordic Optical Telescope (NOT) and the ALFOSC instrument
in UBVRI filters.  The instrument field is $7\arcmin$ on each side so
four pointings were made, covering in total $12\farcm5 \times
12\farcm5$.  The seeing conditions were good and the measured image
qualities in the combined images are 1\farcs0, 1\farcs0, 0\farcs8,
0\farcs6, 0\farcs6 in the UBVRI, respectively.  The field contains
close to 200 galaxies with z$_{spec}$ from Lilly {\it et
al.}~\cite{Lilly95.1} and Koo {\it et al.}~\cite{Koo96.1}.  The
3-$\sigma$ limiting AB magnitudes are $23.7, 25.1, 25.2, 25.5, 25.1$
in the UBVRI, respectively.  Objects were only considered when detected
in a minimum of three bands, this gave a total of $\sim4500$ objects.

\section{Photometric redshifts}

There are two common methods for estimating photometric redshifts.
One method relies on a training sample in which the redshifts are
known and then a regression is made of the observed colours with the
known redshifts.  The second method does not need a training sample,
but instead uses template spectral energy distributions (SEDs),
which are redshifted and fit to the observed colours.

\begin{figure}
\centerline{
  \psfig{figure=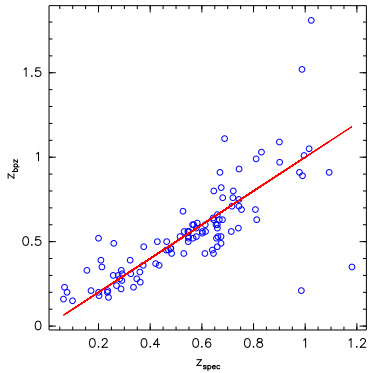,width=0.4\hsize}
}
\caption{Plot showing the estimated photometric redshift versus the spectroscopic redshift for the sub-sample of galaxies for which this information exists}
\label{fig:bpzres}
\end{figure}

We use the BPZ method as implemented by N. Benitez~\cite{Benitez2000.1}.
This method provides an estimate of the redshift, its uncertainty and 
the galaxy type. The results are shown in Fig.~\ref{fig:bpzres}.
At around z$\sim1$ the scatter increases due to the fact that
the main tracer of redshift (the 4000\AA-break) is redshifted out of the 
I-band.

\section{Weak lensing measurements}

\begin{figure}
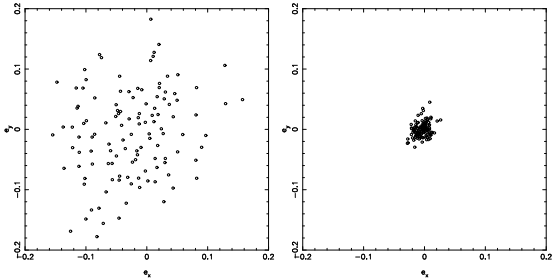

\centerline{
  \psfig{figure=eplot-before-correction.ps,width=0.3\hsize}
  \psfig{figure=eplot-after-correction.ps,width=0.3\hsize}
}
\caption{Plots showing the ellipticity vector components for stars before and after correction for the PSF anisotropy}
\label{fig:psfani}
\end{figure}

The galaxy shapes are measured using N. Kaiser's IMCAT software
package based on the method described in Kaiser {\it et
al.}~\cite{Kaiser95.1}.  Field distortion was found to be non-neglible
due to flexure in the instrument and was corrected by modeling the
distortion as a second order polynomial in each image with respect to
a reference image taken at the USNO 1.0m telescope (thanks to
A. Henden) which is known to have negligible distortion.
PSF anisotropies were also identified in the individual images, but
varying very smoothly from image to image. Again a second order polynomial
model was applied to describe the anisotropy in each image. These models
were then combined in to a final model which was used to correct the
combined image. The success of this correction is shown in 
Fig.~\ref{fig:psfani}.
Finally, seeing effects were normalized by applying the 'pre-seeing
shear polarizability' $P_\gamma = P^{sh} - \frac{P^{sh}(*)}{P^{sm}(*)}
P^{sm}$, giving the corrected $e = e^\prime - \gamma P^\gamma$
(Luppino \& Kaiser~\cite{Luppino97.1}).

\section{Model}

We adopt the simple truncated isothermal sphere model used in
Brainerd {\it et al.}~\cite{Brainerd96.1} 
$$\rho(r) = \frac{V^2 s^2}{4\pi G r^2 (r^2+s^2)}$$
and the scaling relations used in Hudson {\it et al.}~\cite{Hudson98.1}
$$V = V_* \left[ \frac{L}{L_*(z)} \right]^\eta \quad s = s_{200} \left( \frac{V}{200\ {\rm km\ s}^{-1}} \right)^2 .$$
This model is thus used to compute the expected shear for every
source galaxy, taking into account all lens galaxies (see definition of
lens galaxy in next section). The expected shear is then subtracted from the
observed shape of the source galaxy giving the weak lensing
corrected ellipticity, which can be compared to the true ellipticity
distribution. The latter is assumed to be the best fit gaussian to the
observed (raw) ellipticities. 
We hence maximize the corrected ellipticity distribution with respect to
the assumed intrinsic (true) distribution and the likelihood function becomes
$$log L = \sum_i \left( - \frac{| \epsilon_i - \gamma_i P^\gamma_j |}{2 \sigma_e^2} \right) \quad ,$$
where $P^\gamma_j$ is the 'pre-seeing shear polarizability' and $\sigma_e$ is
the best fit Gaussian HWHM. A different shear, $\gamma_i$, is obtained by
varying the parameters of the model (including scaling relations).

\section{Results}

\begin{figure}[t]
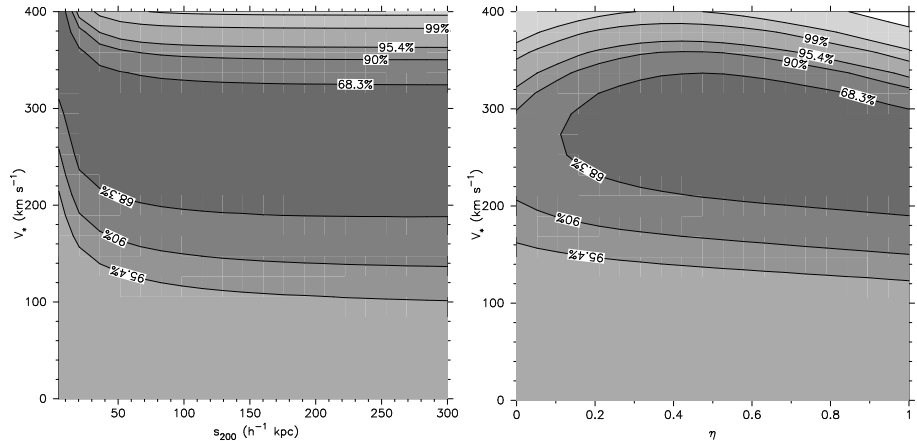

\centerline{
  \psfig{figure=mllh-1.ps,width=0.5\hsize}
  \psfig{figure=mllh-2.ps,width=0.5\hsize}
}
\caption{Two-dimensional weak lensing confidence contours on $V_*$, $s_{200}$ and $V_*$, $\eta$}
\label{fig:wgglres}
\end{figure}

Lenses are selected as having $z < z_s + 0.25$, a maximum $z < 0.5$
and a projected separation in the lens plane in the range $25 < r <
150$ kpc. Furthermore, in order to select secure lenses, a final
criteria is imposed, namely that the photometric redshift has a
probability larger than 95\%.  Due to the fact that the field is
limited and we therefore do not know about the lenses outside the
field-of-view for sources close to the edge, we only consider sources
more than 150 kpc from the image borders at any given lens redshift.
We apply the following cosmological parameters in the analysis:
$\Omega=0.3$, $\Lambda=0.7$ and $h=1$. The results are shown as
confidence contours in Fig.~\ref{fig:wgglres}.

The resulting 68.3\% conf. level on the circular velocity and
the Tully-Fischer exponent is:
$$V_* = 280 \pm 30\ {\rm km\ s}^{-1} \qquad \eta =  0.6 \pm 0.4$$

In the near future with photo-z (optical + infrared photometry) and
large fields one may explore differences in halo mass, size and
scaling relations for different galaxy types and evolution effects.

\end{document}